\documentclass[a4paper,11pt]{article}
\usepackage{pos}
\renewcommand{\logo}{\relax}
\renewcommand{\posurl}{\relax}
\makeatletter
\renewcommand{\PoScopyright@box}{\relax}
\makeatother
\title{Phenomenology of Matching Exponentiated Photonic Radiation to a Parton Shower in KKMChh}
\ShortTitle{Matching KKMChh to a Parton Shower}
\author*[a]{S.A. Yost}
\author[b]{B.F.L. Ward}
\author[c]{Z. Was}
\affiliation[a]{Physics Department, The Citadel,\\
  171 Moultrie St., Charleston, SC 29409, U.S.A.}
\affiliation[b]{Physics Department, Baylor University,\\
1311 S 5th St., Waco, TX 76706, U.S.A.}
\affiliation[c]{Institute of Nuclear Physics Polish Academy of Science,\\
ul. Radzikowskiego 152 31-342 Krak{\'o}w, Poland}
\emailAdd{scott.yost@citadel.edu}
\emailAdd{bfl\_ward@baylor.edu}
\emailAdd{zbigniew.was@ifj.edu.pl}
\abstract{KKMChh adapts the soft photon exponentiation of the program KKMC, initially for electron-positron annihilation, to hadron collisions, where it must interface to 
a parton shower and parton distribution functions (PDFs) which may already include effects of QED radiation. We describe the NISR (Negative Initial State Radiation) 
algorithm  developed to consistently interface with PDFs including QED effects, and present results on its effect on some distributions of phenomenological interest.}
\FullConference{17th International Symposium on Radiative Corrections: Applications of Quantum Field Theory to Phenomenology (RADCOR2025)\\
5-10 October 2025\\
Puri, India\\}
\begin{document}
\maketitle

\section{Introduction}
KKMC~\cite{Jadach:1999vf} was developed for Precision Z boson phenomenology in collisions of the form 
$e^+ e^- \rightarrow Z/\gamma^* \rightarrow f{\overline f} + n\gamma$ including 
exact $O(\alpha)$ and $O(\alpha^2 L)$ QED initial state radiation (ISR), final state 
radiation (FSR) and initial-final interference (IFI) with soft photon
exponentiation implemented at the amplitude level in the CEEX 
formalism~\cite{Jadach:1998jb,Jadach:2000ir}. The original
KKMC program has been upgraded and released as KKMCee 5.00.00~\cite{KKMCee:2023}.

The program KKMChh~\cite{KKMChh:2016,KKMChh:2019} implements the same amplitude-level photon resummation for 
hadronic collisions, $pp\rightarrow Z/\gamma^* \rightarrow l{\overline l} + n\gamma$.
We focus here on the implementation of QED ISR in KKMChh and its matching to parton distribution functions (PDFs) which may include their own QED ISR effects. 

At a simple level, the total Drell-Yan cross section may be expressed as 
\begin{eqnarray}
\sigma(s) &=& \frac{3\pi}{4}\sigma_0(s) \sum_{q\in\{d,u,s,c,b\}} \int d{\hat x} dz 
\int dx_q dx_{\bar q} \delta({\hat x} - x_q x_{\bar q})f_q(x_q, {\hat x}s)
 f_{\bar q}(x_{\bar q}, {\hat x}s) \nonumber\\
& & \qquad \qquad \qquad \times \;\rho_{\rm ISR}^{(0)}(1 - z, {\hat x}s) \sigma^{\rm Born}_{q{\bar q}}(z{\hat x}s)
\langle W_{\rm MC}\rangle
\end{eqnarray}
where $f_q, f_{\bar q}$ are the PDFs for the quark and antiquark, $\rho_{\rm ISR}$ is 
the YFS ISR radiator function~\cite{yfs}, ${\hat x}s = (p_q + p_{\bar q})^2$ is the squared
momentum of the quarks, and $z = 1-v$ where $v$ is the total ISR energy fraction
defined via the relation
\begin{equation}
\sum_{\gamma\in{\rm ISR}} E_\gamma = v\sqrt{{\hat x}s}.
\end{equation}
By design, the MC weights are clustered close to 1 and to leading order, 
\begin{eqnarray}
\rho_{{}_{\scriptstyle\rm ISR}} &=& F_{\rm YFS}(\gamma)\gamma v^{\gamma-1}\ ,\label{eq3} \\
F_{\rm YFS}(\gamma) &=& \frac{e^{-C_E \gamma}}{\Gamma(1+\gamma)}\ , \\
\gamma = \gamma_{{}_{\scriptstyle\rm ISR}}({\hat x}s, Q_q, m_q) & = & \frac{2\alpha}{\pi} Q_q^2
\left[\ln\left(\frac{{\hat x}s}{m_q^2}\right) - 1\right].\label{eq5}
\end{eqnarray}
The logarithm in Eq.~(\ref{eq5}) is the the large logarithm $L$ in the ISR 
leading-logarithm expansion.
KKMChh generates $x_q, x_{\bar q}, v$ and the quark flavor $q$ using the adaptive 
Monte Carlo generator FOAM.~\cite{Jadach:2002kn}

The main focus of this paper is the interaction of KKMChh's exponentiated photonic ISR with the parton distribution function, especially in for use with PDF sets that already account for QED explicitly. Even when the QED content is not explicit, there can be  QED contamination in the input data. QED-corrected parton distribution sets are not ideal on their own, because the transverse momentum component of photonic ISR can be significant and cannot be modeled by a collinear PDF to the precision of the ISR 
calculations in KKMChh.  
To allow KKMChh to work with current PDF sets including QED effects, we introduced a feature “Negative ISR” (NISR) to back out the QED component starting at a given scale and then generating ISR via KKMC’s algorithm.
We begin by describing the effect of ISR on the parton luminosity distribution, and then describe the NISR mechanism and its effects. Finally, we show a phenomenological
application of NISR in the context of forward-backward asymmetry studies.

\section{The Effect of ISR on Parton Luminosity}

To understand the effect of ISR, we can focus on the joint parton 
luminosity function for each quark flavor, 
\begin{equation}
xL_{q{\bar q}}(xs) \equiv \int dx_q dx_{\bar q}
\;\delta(x - x_q x_{\bar q})\; x_q f_q(x_q, xs)\;
 x_{\bar q} f_{\bar q}(x_{\bar q}, xs)\ .
\end{equation}
Including the QED ISR radiator function defines a QED-corrected joint parton luminosity function 
\begin{eqnarray}
xL_{q{\bar q}}^{\rm QED}(xs) &\equiv& \int dz d{\hat x}\; \delta(x - z{\hat x})\; {\hat x} L_{q{\bar q}}({\hat x}s) \; z \rho_{{}_{\scriptstyle\rm ISR}}(1 - z, {\hat x}s) \nonumber\\ 
&=& \int_0^{1-x} dv\; \delta\left({\hat x} - \frac{x}{1-v}\right)\; {\hat x} L_{q{\bar q}}({\hat x}s)\;\rho_{{}_{\scriptstyle\rm ISR}}(v, {\hat x}s)\ .
\label{eq6}
\end{eqnarray}

\begin{figure}[!ht]
    \includegraphics[width=0.45\textwidth]{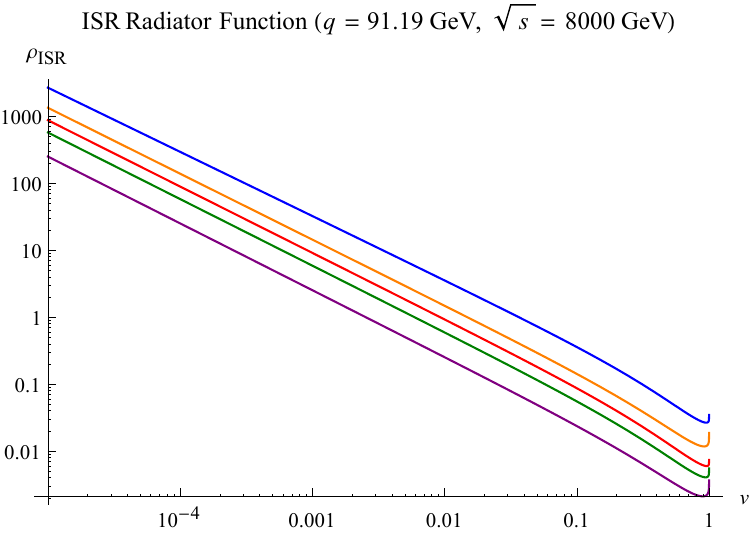}
    \includegraphics[width=0.53\textwidth]{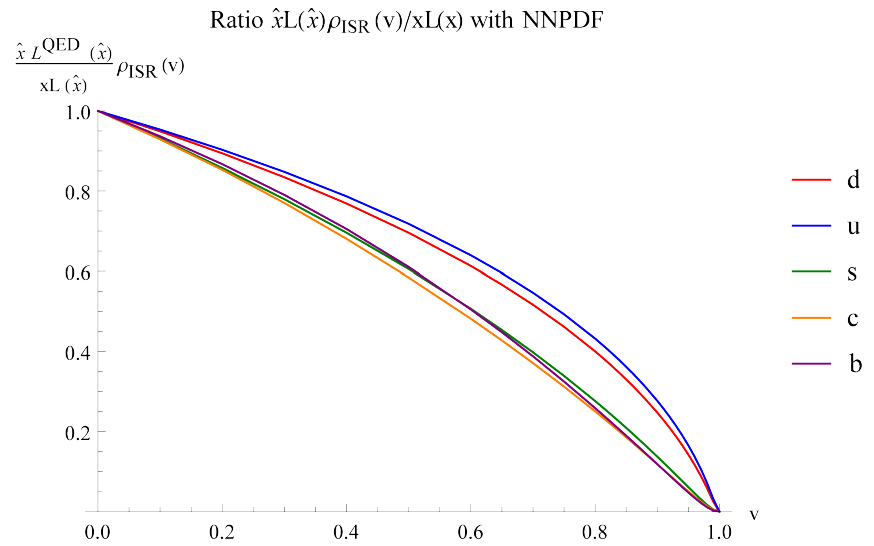}
    \caption{The left graph shows the QED ISR radiator function $\rho_{\rm ISR}(v)$ as a function of the ISR photon energy fraction $v$, and the left graph shows the ratio of the integrand of Eq.~(\ref{eq6}) to $xL_{q{\bar q}}(xs)$ without QED corrections}\label{fig1}
\end{figure}

Fig.~\ref{fig1} shows the ISR radiator function $\rho_{\rm ISR}$ as a function of the 
ISR energy fraction $v$ on the left, and the ratio the integrand of Eq.~(\ref{eq6}) to
$L_{q{\bar q}}(xs)$ without QED corrections on the right.

\begin{figure}[!ht]
    \includegraphics[width=0.49\textwidth]{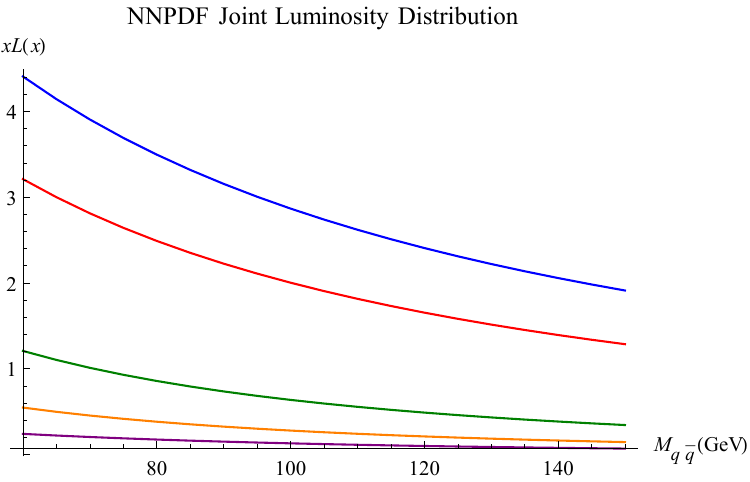}
    \includegraphics[width=0.49\textwidth]{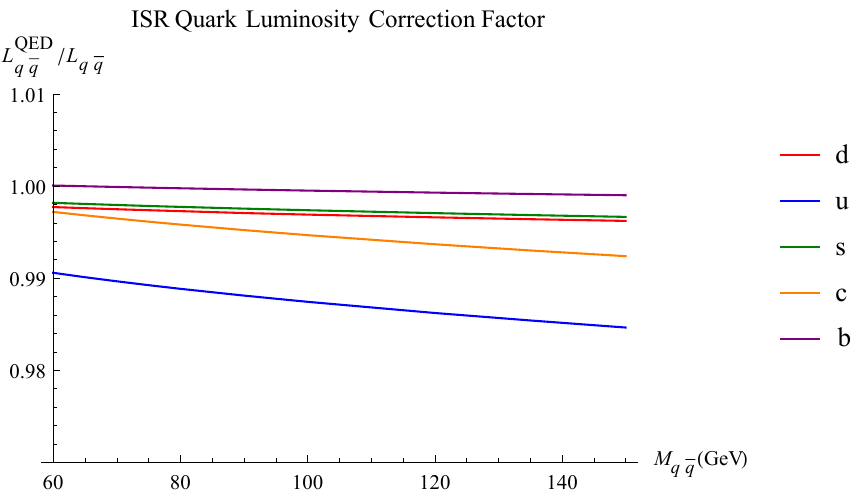}
    \caption{The QED-corrected joint parton luminosity functions 
$x L^{\rm QED}_{q{\bar q}}$ for each flavor on the left, together with the
ratio to the original NNPDF joint parton luminosity function on the right, 
for $M_{q{\bar q}}$ between 60 and 140 GeV.}\label{fig2}
\end{figure}

\begin{figure}[!ht]
    \includegraphics[width=0.49\textwidth]{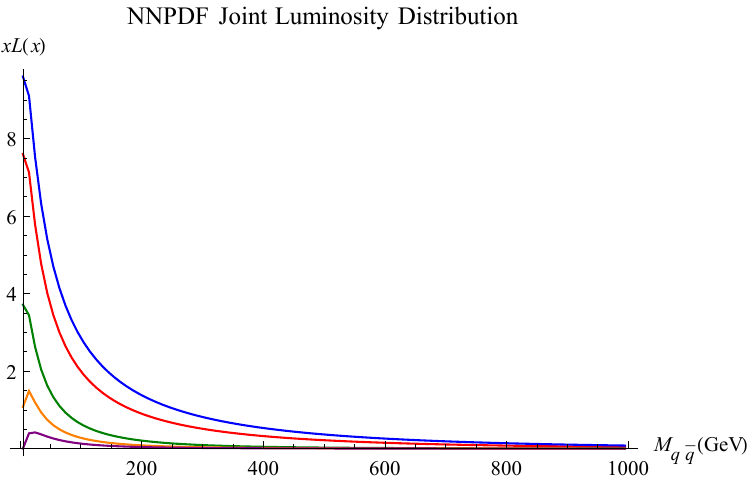}
    \includegraphics[width=0.49\textwidth]{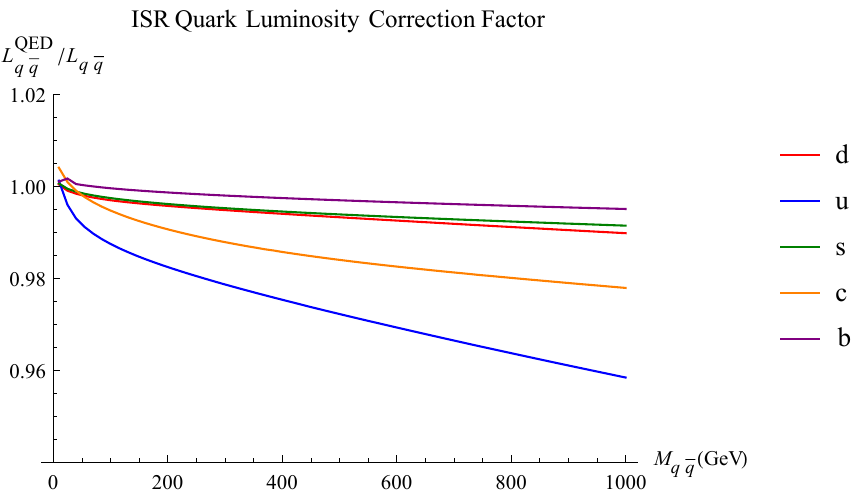}
    \caption{The QED-corrected joint parton luminosity functions 
$x L^{\rm QED}_{q{\bar q}}$ for each flavor on the left, together with the
ratio to the original NNPDF joint parton luminosity function on the right, 
for $M_{q{\bar q}}$ up to 1000 GeV.}\label{fig3}
\end{figure}

Figs.~\ref{fig2} and \ref{fig3} show the QED-corrected joint parton luminosity functions on the left together with their ratios to the original functions on the right for NNPDF3.1 NLO~\cite{nnpdf1} parton distribution functions.\footnote{The choice of parton distribution function has little effect on the QED corrections shown.}
Fig.~\ref{fig2} shows a 
range of $M_{q{\bar q}}$ between 60 and 140 GeV and Fig.~\ref{fig3}  shows a 
range up to 1000 GeV. All figures were generated using ManeParse \cite{ManeParse}. We use the current quark masses 
\begin{equation}
m_u = 2.2\ {\rm MeV},\ m_d = 4.7\ {\rm MeV},\  m_s = 150\ {\rm MeV},\  m_c = 1.2 \ {\rm GeV},\ 
m_b = 4.6\  {\rm GeV}.
\end{equation}

\begin{figure}[!ht]
    \includegraphics[width=0.47\textwidth]{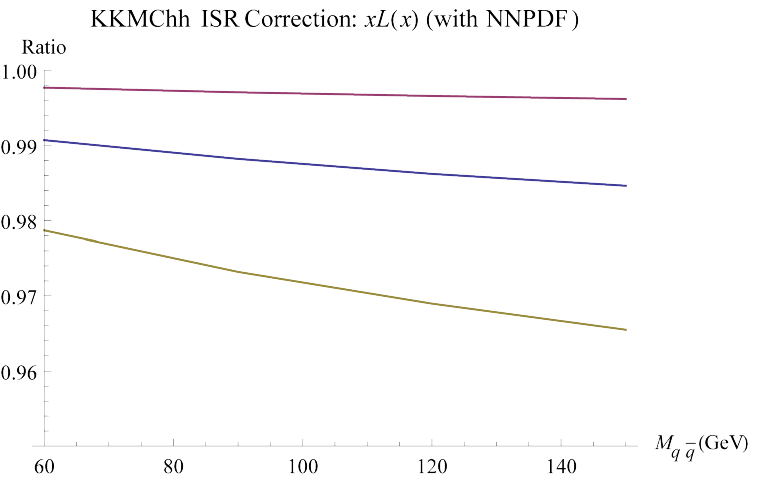}
    \includegraphics[width=0.51\textwidth]{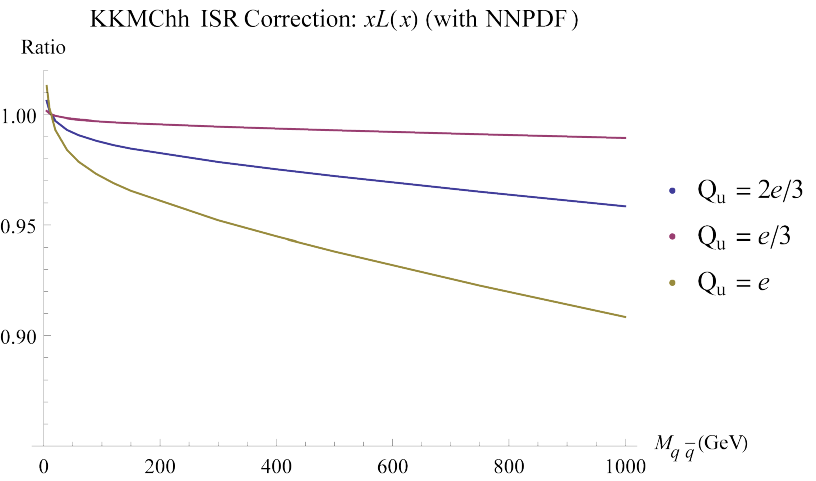}
    \caption{The ratio of the up quark QED-corrected joint parton luminosity function 
$x L^{\rm QED}_{q{\bar q}}$ to the original NNPDF joint parton luminosity function
with the charge varied from $2e/3$ to $e/3$ to $e$ for energy scales between
60 and 140 GeV on the left and up to 1000 GeV on the right.}\label{fig4}
\end{figure}

\begin{figure}[ht]
    \includegraphics[width=0.42\textwidth]{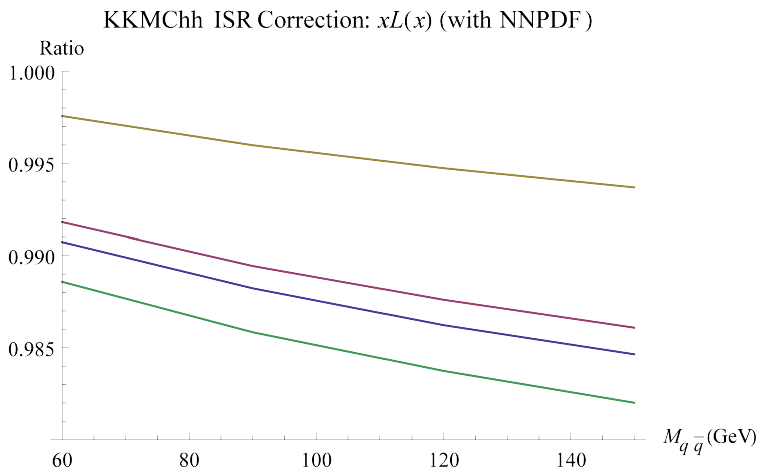}
    \includegraphics[width=0.56\textwidth]{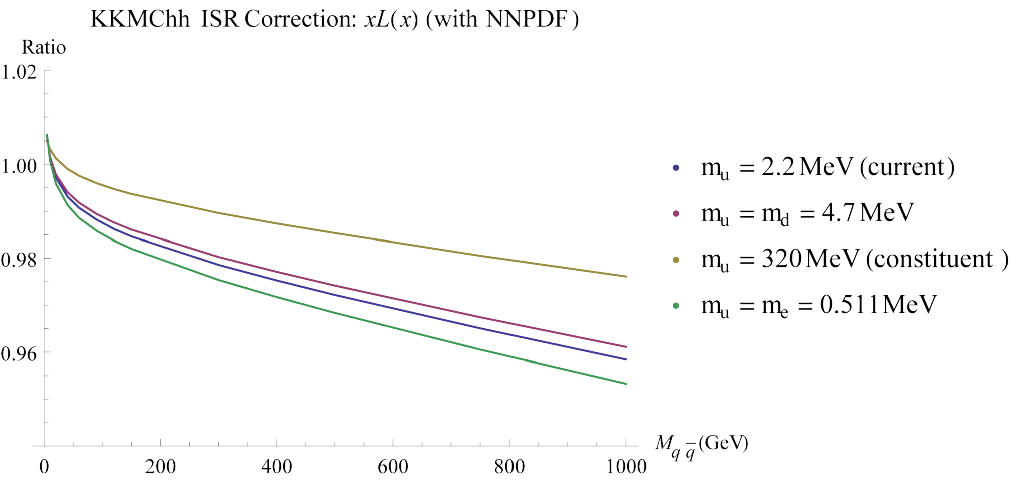}
    \caption{The ratio of the up quark QED-corrected joint parton luminosity function 
$x L^{\rm QED}_{q{\bar q}}$ to the original NNPDF joint parton luminosity function
with the mass varied from $m_u = 2.2$ MeV to 4.7 MeV, 320 MeV, and 0.511 MeV, 
for energy scales between 60 and 140 GeV on the left and up to 1000 GeV on the right.
}\label{fig5}
\end{figure}

The dominance of up-quark photon radiation is explained mostly by its charge,
with the smallness of its mass having a lesser influence, as illustrated in 
the subsequent figures. Fig.~\ref{fig4} varies the up quark charge from 
$2e/3$ to $e/3$ and $e$ to isolate the effect of the quark charge while leaving
the mass fixed. The up quark QED correction factor is almost identical to the 
down quark factor when its charge is changed from $2e/3$ to $e/3$ without 
changing the mass. Fig.~\ref{fig5} replaces the up quark's current mass 
$m_u = 2.2$ MeV with three alternatives: the down quark mass $m_d = 4.7$ MeV,
the up quark's constituent mass 320 MeV, and the electron mass $m_e = 0.511$
MeV. The mass dependence, being logarithmic in $\gamma_{\rm ISR}$, is much
weaker than the charge dependence. 

\section{Matching Exponentiated QED to Parton Distribution Functions}
A number of parton distribution functions are available that combine QCD and 
QED evolution and include a photon PDF. The PDF model uses a collinear 
approximation for the photon distribution. This is considerably less accurate
than KKMChh's treatment of ISR, which includes complete $O(\alpha)$ and 
up to next-to-leading logarithm $O(\alpha^2 L)$ contributions, with soft 
photons resummed to all orders. Thus, it is desirable to match KKMChh's 
calculations onto a PDF without double-counting photon ISR. Even when a PDF
set does not have QED evolution, the input data can contain QED contamination 
that could lead to some double-counting of ISR. 

We would like to have a PDF set that represents the full data, modeled with QCD and QED together, below the hard process scale $Q_0$ and prunes out the QED component for $Q>Q_0$, where KKMChh will model the ISR QED. To allow KKMChh to work with current PDF sets including QED effects, we introduced a feature ``Negative ISR,'' or NISR, to back out the QED component starting at a scale $Q_0$ and then generate ISR via KKMC’s more sophisticated algorithm.

The pruning of ISR from the PDFs would ideally be done in advance, but can be done on
the fly, allowing standard PDF sets to be used for event generation. The net effect
is to convolute each quark PDF with reversed ``half-radiator'' functions 
\begin{equation}
\rho_{\rm ISR} \left(-\frac12 \gamma_{\rm ISR}(xs), u_1\right), \qquad
\rho_{\rm ISR} \left(-\frac12 \gamma_{\rm ISR}(xs), u_2\right)
\end{equation}
whose convolution makes a full reverse radiator function.\footnote{In practice, the reverse radiators are calculated including leading hard and virtual corrections to Eq.~(\ref{eq3}).} This introduces two new variables $u_1$ and $u_2$
into the primary KKMChh distribution, subject to the constraint $x = {\hat x}(1-u_1)
(1-u_2)$. The modified ISR energy fraction becomes $v'$, where 
\begin{equation}
1-v'= (1-v)(1-u_1)(1-u_2)
\end{equation}
and the quark momentum fractions become 
\begin{equation}
x_1' = x_1 (1-u_1)\ , \qquad x_2' = x_2(1-u_2).
\end{equation}
There are now six variables in the primary initial state MC distribution, generating
the quark flavor and the variables ${\hat x}, x_1, v, u_1$, and $u_2$. 

When used with a PDF set incorporating QED corrections, 
the NISR scale should be set at the hard process scale.  To remove QED contamination from a PDF set without QED evolution, it would be more appropriate to 
set a low  NISR scale, at the starting point of QCD evolution, with $Q_0$ typically on the order of few GeV. We have run tests with NNPDF3.1-LUXqed~\cite{nnpdf2a,nnpdf2b,nnpdf2c} and a variety of other PDF sets with and without QED corrections, and expect to report on those tests elsewhere. 

The implementation of NISR can be checked by comparing the cross section obtained with a given PDF set alone to the cross section obtained using the PDF set with KKMChh’s ISR and NISR together. In these tests, NISR is applied at the hard process scale, which 
would be appropriate with a PDF set incorporating QED corrections.\footnote{NNPDF3.1 NLO
PDFs were used for these tests, but the 
underlying PDF set has little effect on the
results.}
We can also check the independence on the quark mass, since the logarithmic dependence should cancel through second order. 
The first two columns in table \ref{tab1} compare the Born cross section to a QED-corrected one including NISR, while the third column shows the difference, which should be due to a NLO non-logarithmic correction 
$\delta_Q = Q^2 \frac{\alpha}{\pi}\left(-\frac12 + \frac{\pi^2}{3}\right),$ which 
is $0.3\%$ for up-type quarks and $0.08\%$ for down-type quarks. This is consistent
with the results in the table.

\begin{table}[!ht]
\centering
\begin{tabular}{|l|c|c|c|}
\hline
\multicolumn{4}{|c|}{Combined ISR + NISR effect with Current Quark Masses}\\
\hline
Quark & Born (No Photons) & QED: ISR + NISR (no FSR) & Fractional Difference \\
\hline 
Down 	& 365.55 ± 0.00 & 365.86 ± 0.02 & 0.09 ± 0.01 \\
Up 	& 420.68 ± 0.01 & 421.43 ± 0.05 & 0.18 ± 0.01 \\
Strange & 120.66 ± 0.00 & 120.79 ± 0.01 & 0.11 ± 0.01 \\
Charm 	& 43.12 ± 0.00 	& 43.24 ± 0.01 	& 0.27 ± 0.02 \\
Bottom 	& 24.23 ± 0.00 	& 24.23 ± 0.00 	& 0.07 ± 0.02 \\
\hline
Total 	& 974.23 ± 0.01 & 975.55 ± 0.05 & 0.14 ± 0.01 \\
\hline
\end{tabular}
\caption{
Comparison of the Born cross section without photons to the QED-corrected Born 
cross section with ISR and NISR together, but no FSR. The last row shows the five-quark
total cross section. 
}
\label{tab1}
\end{table}

\begin{table}[!ht]
\centering
\begin{tabular}{|l|c|c|c|}
\hline
\multicolumn{4}{|c|}{Combined ISR + NISR effect with Quark Masses Replaced by 500 MeV}\\
\hline
Quark & Born (No Photons) & QED: ISR + NISR (no FSR) & Fractional Difference \\
\hline 
Down 	& 365.55 ± 0.00 & 365.81 ± 0.02   & 0.07 ± 0.01 \\
Up 	& 420.68 ± 0.01 & 421.43 ± 0.05   & 0.24 ± 0.00 \\
Strange & 120.66 ± 0.00 & 120.79 ± 0.01   & 0.08 ± 0.00 \\
Charm 	& 43.12 ± 0.00  & 43.24 ± 0.01    & 0.27 ± 0.01 \\
Bottom 	& 24.23 ± 0.00  & 24.23 ± 0.00    & 0.08 ± 0.01 \\
\hline
Total 	& 974.23 ± 0.0  & 1,975.55 ± 0.05 & 0.16 ± 0.0 \\
\hline
\end{tabular}
\caption{
Comparison of the Born cross section without photons to the QED-corrected Born 
cross section with ISR and NISR together, but no FSR with all quark masses replaced 
by 500 MeV to verify the quark-mass independence of the results.  
}
\label{tab2}
\end{table}

The logarithmic dependence of ISR on the quark masses should cancel between ISR and 
NISR through second order. Table \ref{tab2} verifies this by replacing all of the 
quark masses by 500 MeV and repeating the comparisons of Table \ref{tab1}. The 
net effect of ISR + NISR on the Born cross section is again found to be almost 
exactly the same, in line with expectations. 

\section{Application to Forward-Backward Asymmetry Studies}

KKMChh has been applied to studies of forward-backward asymmetry, where initial-final
interference is important and KKMChh has the unique advantage of an exponentiated $O(\alpha^2L)$ calculation of this effect.~\cite{kkmchh-afb,kkmchh-afb-arxiv} Here, we
examine the effect of applying NISR to the previous calculations. 

\begin{figure}[!ht]
\begin{center}
    \includegraphics[width=\textwidth]{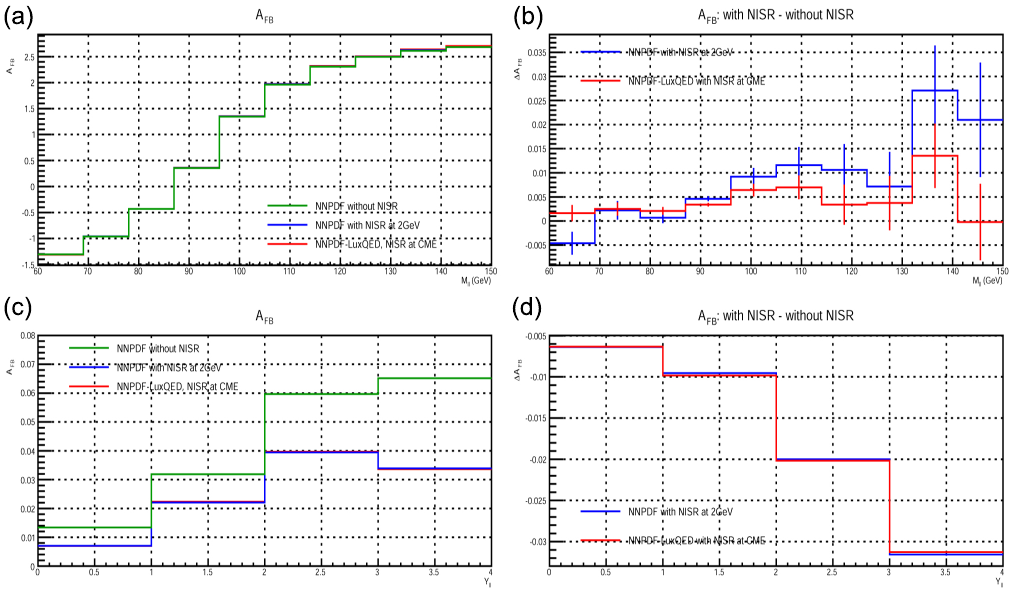}
    \caption{Plots (a) and (c) show $A_{\rm FB}$ as a function of the muon pair invariant mass $M_{\mu^-\mu^+}$ and rapidity $Y_{\mu^-\mu^+}$, for the three cases (1) NNPDF3.1 without NISR in green, (2) NNPDF3.1 with NISR at 2 GeV in blue, and (3) NNPDF3.1-LUXqed with NISR at the hard process scale in red. Plots (b) and (d) show the corresponding differences $(2) - (1)$ in blue and $(3) - (1)$ in red. }\label{fig6}
\end{center}
\end{figure}

In these tests, 
KKMChh generated muon final states in proton-proton collisions at an 8 TeV CM energy.
No hadronic shower was applied to these events, and the 
muon momenta are unconstrained. All events include final-state radiation (FSR), while both initial-state
radiation (ISR) and initial-final interference (IFI) can be switched, allowing these
effects to be examined independently. The results in Fig.~\ref{fig6} are obtained from
a sample of $1.04\times 10^{10}$ events using NNPDF3.1 NLO PDFs with NISR applied at 2
GeV, and using NNPDF3.1-LUXqed NLO PDFs with NISR applied at the hard process scale. 

Plot (a) in Fig.\ \ref{fig6} shows the forward-backward asymmetry $A_{\rm FB}$ as a 
function of the muon pair invariant mass $M_{\mu^-\mu^+}$ for three cases: (1) NNPDF3.1
without NISR in green, (2) NNPDF3.1 with NISR at 2 GeV in blue, and (3) NNPDF-LUXqed with NISR at the hard process scale in red. In plot (b),
the difference $(2) - (1)$ is shown in blue and the difference $(3) - (1)$ is shown
in red. Plots (c) and (d) make the same comparisons as a function of the muon pair 
rapidity $Y_{\mu^-\mu^+}$. The effect of NISR is small for invariant masses up to 130 GeV, and may increase above 130 GeV, but the statistics in this region are low. The 
effect of NISR increases with rapidity, but is largely independent of whether it is 
applied at the hard process scale with the LUXqed PDF set or applied at 2 GeV with the
ordinary PDF set. 

\section{Conclusion}
To avoid double-counting of initial state QED radiation, ISR can be backed out of 
a PDF set using the NISR procedure outlined in the previous section before adding  
the ISR generated by KKMChh. 
We expect to report soon on further tests applying NISR to calculations of 
forward-backward asymmetry. We have reported on some phenomenological effects earlier,~\cite{ichep2022,ichep2024} and a more detailed paper on the formalism and phenomenology of NISR should appear soon. In many 
cases, the effect of NISR is very minimal. When it is below the desired precision level of a calculation, it would be better to omit it, since the NISR generation adds 
significant time to a MC run. Generating PDF sets ``pruned'' of QED ISR in advance 
would make this process more efficient.  

\acknowledgments

We acknowledge the essential contributions of Stanisław Jadach (1947 -- 2023), a creator of KKMC and leader in the field of precision radiative corrections.
S.A.\ Yost was supported in part by a grant from the Citadel Foundation. Z. Was was supported in part by Narodowe Centrum Nauki, Poland, grant No. 2023/50/A/ST2/00224.

\end{document}